\documentclass[10pt, a4paper]{article}

\usepackage[T1]{fontenc}
\usepackage{lmodern}

\usepackage{microtype}

\usepackage{hyperref}
\hypersetup{
    colorlinks=true,
    allcolors=black,
    breaklinks=true,
    }

\usepackage[
    style=apa,
    sortcites=true,
    sorting=nyt,
    backend=biber,
    autocite=inline
    ]{biblatex}

\usepackage{csquotes, xpatch}

\usepackage{graphicx}
\graphicspath{ {figs} {}}

\usepackage{amsmath}

\usepackage[font=footnotesize, labelfont=it, format=plain, width=.9\textwidth]{caption}
\usepackage{subfig}

\usepackage{wrapfig}
\usepackage{listings}
\usepackage{lstbayes}

\usepackage{xcolor}

\definecolor{codegreen}{rgb}{0,0.6,0}
\definecolor{codegray}{rgb}{0.5,0.5,0.5}
\definecolor{codepurple}{rgb}{0.58,0,0.82}
\definecolor{backcolour}{rgb}{0.96,0.96,0.96}

\usepackage[scale=0.9]{sourcecodepro}
\lstdefinestyle{mystyle}{
    backgroundcolor=\color{backcolour},
    commentstyle=\color{codegreen},
    numberstyle=\tiny\color{codegray},
    basicstyle=\ttfamily\footnotesize,
    breakatwhitespace=false,
    breaklines=true,
    captionpos=b,
    keepspaces=true,
    numbers=none,
    numbersep=5pt,
    showspaces=false,
    showstringspaces=false,
    showtabs=true,
    tabsize=2,
    columns=fullflexible,
}

\usepackage[
    firstpageonly=false,
    angle=25,
    text=Draft,
    fontsize=24pt,
    pos={2.5cm,2.5cm},
    anchor=cm,
    stamp=false
    ]{draftwatermark}

\usepackage{enumitem}
\setlist{noitemsep}

\usepackage{booktabs}
\usepackage[pagewise]{lineno}

\title{Method for recovering data on unreported low-severity crashes}
\author{Alberto Morando \thanks{Autoliv Development AB; \url{alberto.morando@autoliv.com}}}
\date{March 6, 2025}

\begin{document}
\maketitle

\begin{abstract}
    \textbf{Objective:} Many low-severity crashes are not reported due to sampling criteria, introducing missing not at random (MNAR) bias. If not addressed, MNAR bias can lead to inaccurate safety analyses. This paper illustrates a statistical method to address such bias.
    \textbf{Methods:} We defined a custom probability distribution for the observed data as a product of an exponential population distribution and a logistic reporting function. We used modern Bayesian probabilistic programming techniques.
    \textbf{Results:} Using simulated data, we verified the correctness of the procedure. Applying it to real crash data, we estimated the $\Delta v$ distribution for passenger vehicles involved in personal damage-only (PDO) rear-end crashes. We found that about 77\% of cases are unreported.
    \textbf{Conclusions:} The method preserves the original data and it accounts well for uncertainty from both modeling assumptions and input data. It can improve safety assessments and it applies broadly to other MNAR cases.

    \vspace{5mm}
    \noindent\footnotesize\textbf{Keywords: } Traffic safety; Crash modelling; Safety impact; Safety benchmark; Data Imputation.

    \vspace{3mm}
    \noindent\footnotesize\textbf{Word count: } 3397 

\end{abstract}

\section{Introduction}
Sampling bias, when the manner of data collection systematically excludes certain portions of the population, is common in observational research. Left uncorrected, this bias can lead to wrong inferences, because the observations do not accurately represent the true distribution of the data. The bias needs to be understood, modeled, and corrected for the specific situation at hand to ensure accurate results \autocite{allisonMissingData2009}. Systematic bias that results in missing not at random (MNAR) data is the most challenging; this bias is not immediately clear from the data alone, as it depends on unobserved factors and the observed data.

Crash databases are susceptible to MNAR data. For example, a crash reported in the \textit{Crash Investigation Sampling System} (CISS) must involve at least one passenger vehicle towed away from the scene \autocite{nhtsa-ciss,Radja2022}. Because of this sampling criterion, the CISS is biased towards high-severity crashes; many crashes with uninjured occupants are MNAR. As a result, injury estimates (and similar analyses) of crash databases with MNAR data, like CISS, are likely to produce inaccurate results (\cite{andricevicInjuryRiskFunctions2018,lubbeInjuryRiskCurves2024}; \cite[][Ch. 2]{blincoe2023economic}).

Constructing an accurate model in the presence of MNAR data requires domain expertise and detailed knowledge of the data collection process. However, while the criteria for reporting a crash might be known, translating them into statistical terms may not be easy. The method presented in this paper uses a Bayesian framework to address this issue. This type of framework is well-suited for constructing statistical models from first principles of probability \autocite{kruschkeBayesianDataAnalysis2017}: Bayesian models maintain transparency in their structure and retain the inherent uncertainty in estimation (from modeling assumptions and missing data).

\section{Example with simulated data} \label{sec-example-simulated}
\subsection{Data}
\begin{figure}[tb]
    \includegraphics[scale=0.9, keepaspectratio]{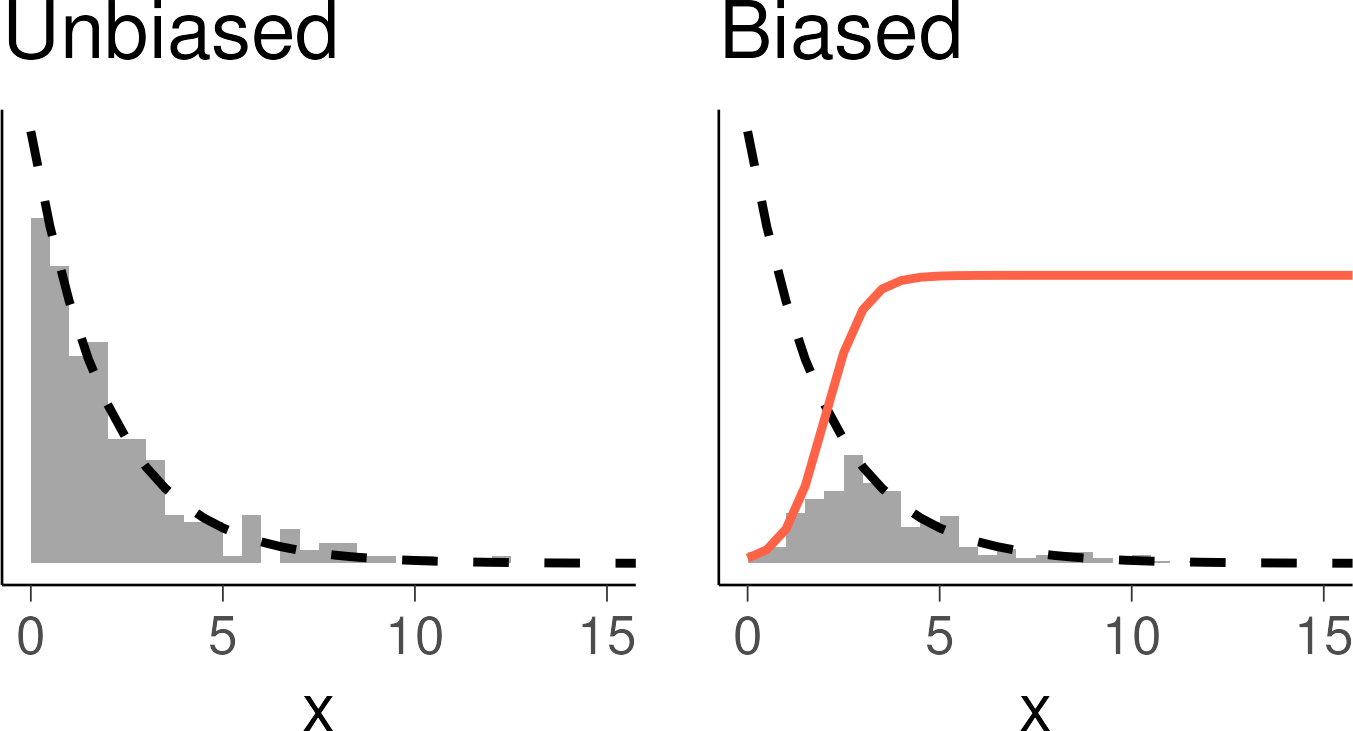}
        \centering
        \caption{Unbiased and biased samples (gray) from the same population (dashed lines). The biased sample originated from a selection mechanism based on the logistic function (red line).}
        \label{fig-nobias-vs-bias}
\end{figure}

In this example, the correctness of the procedure is demonstrated with simulated data. Assume that the probability distribution function (PDF) of the population is exponential, with $\lambda = 0.5$ \autocite{mcstanUnboundedContinuousExponential}. Ideally, a sample $X$ from the population would mimic the underlying distribution accurately (Fig.~\ref{fig-nobias-vs-bias}):

\begin{equation}
    p_{\text{unbiased}}(x \mid \lambda) = \text{Exp}(x \mid \lambda)
    \label{eq-p-x-unbias}
\end{equation}

However, the sampling process is biased and yields a set of observations  $x_{i \dots N}$ ($N_\text{obs} = 250$) that provides only partial information on the population (Fig.~\ref{fig-nobias-vs-bias}). Assume that the probability of observing $x_i$ is weighted according to the cumulative distribution function (CDF) of the logistic distribution (also known as the logistic function), $\mu = 2$ and $\sigma = 0.5$ \autocite{mcstanUnboundedContinuousLogistic}. The logistic function is commonly used to map input variables to a probability outcome. This sampling process systematically undersamples the left tail of the distribution. The unnormalized probability of an observation is an exponential-logistic function:

\begin{align}
    p_\text{biased}(x \mid \lambda, \mu, \sigma) & \propto \text{Exp}(x \mid \lambda) \, \int_{0}^{\infty} \text{Logistic}(x \mid \mu, \sigma) \, dx \\
        & \propto \lambda e^{-\lambda x} \, \frac {1}{1+e^{-(x-\mu )/\sigma}}
    \label{eq-p-x}
\end{align}

where Eq.~\ref{eq-p-x} omits the normalization constant that makes the PDF integrate to 1.

The next step is to create a model for Eq.~(\ref{eq-p-x}) so we can complement the biased set of observations with new data coming directly from the exponential distribution (Eq.~\ref{eq-p-x-unbias}).

\subsection{Method} \label{sec-fake-model} 
We defined a computational statistical model to infer the parameter values of Eq.~(\ref{eq-p-x}) from a set of observations. The Bayesian framework was selected because it has a comprehensive and intuitive account of uncertainties \autocite{kruschkeBayesianDataAnalysis2017} . The model was built \textit{R} (v. 4.2.3; \cite{R}) and the package \textit{brms} (v. 2.22.5; \cite{burknerBrmsPackageBayesian2017}), which is an interface for the probabilistic programming language \textit{Stan} (v. 2.32.2, \cite{carpenterStanProbabilisticProgramming2017}) via the \textit{CmdStanR} backend (v. 0.8.1; \cite{cmdstanr}). The package \textit{brms} makes working with Bayesian models easy. To speed up inference, we used the Variational Inference (VI) algorithm \textit{pathfinder} \autocite{pathfinder2022} instead of a Markov Chain Monte Carlo (MCMC) sampler. While VI is generally less precise than MCMC, new methods like pathfinder work well—and they are much faster.

Although brms and Stan include many common probability functions, there is no specific exponential-logistic function for Eq.~\ref{eq-p-x}. It is possible, however, to define complex custom distributions from simpler probability statements. While the steps can appear intricate and daunting at first, most of the code is boilerplate and can be adapted to the problem at hand with only minor modifications. Details are in the Appendix and the code repository.

Everything is known about the simulated data, including the underlying distributions and the sampling process (Eq.~\ref{eq-p-x}). In contrast, when real data are being analyzed, this \textit{method} section would contain all the intuitions, hypotheses, and prior knowledge that justify the modeling choices.

Finally, Bayesian models require priors. In this case, we placed vague priors to regularize the model and prevent sampling errors:

\begin{equation}
    \lambda, \, \mu, \, \sigma  \sim \mathcal{N}(\mu = 0, \, \sigma = 10)
\end{equation}

and this equation represents another opportunity to include prior information, especially if the data are sparse \autocite{kruschkeBayesianDataAnalysis2017}.

\subsection{Resampling} \label{sec-fake-resample}

\begin{figure}[tb]
    \includegraphics[scale=0.9, keepaspectratio]{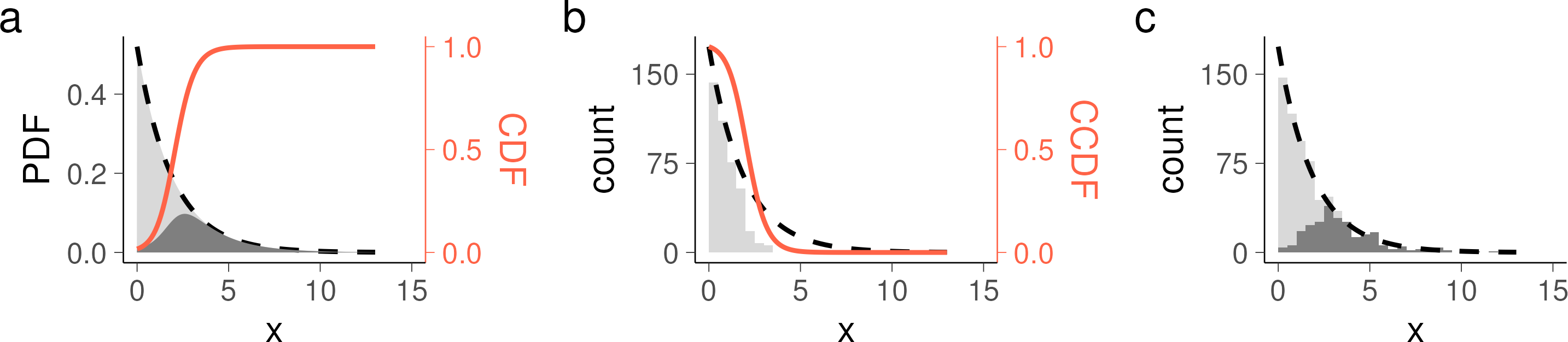}
        \centering
        \caption{(a) The new sample size is proportional to the observed sample size based on the ratio of the population's probability density function (PDF; light gray) to the unnormalized exponential-logistic distribution (dark gray), which results from multiplying the exponential PDF (dashed black line) by the CDF of the logistic distribution (red line). (b) A new set of observations $\text{N}_\text{new}$ is sampled from the exponential PDF (black dashed line) weighted by the complementary CDF (CCDF) of the logistic distribution (red line) following an inverse probability weighting. (c) The union of the new (light gray) and the old (dark gray) sets of observations form an unbiased sample that mimics the exponential distribution of the population (black dashed line).}
        \label{fig-resampling}
\end{figure}

From the model, we retrieved an unbiased sample while maintaining the original sample intact. The procedure is based on an inverse probability weighting (Fig.~\ref{fig-resampling}). Each data point from the population distribution was assigned a weight in proportion to the \textit{inverse} of its probability of being sampled under the biased process. That is, we used the complementary CDF of the logistic distribution [CCDF] to recover data that were neglected with the logistic-based sampling process. The number of new samples $\text{N}_\text{new}$ required to complete the dataset is:

\begin{align}
    \text{N} &= \text{N}_\text{obs} + \text{N}_\text{new} \label{eq-n_a}\\
    \text{N} &= k\text{N}_\text{obs} \label{eq-n_b}
\end{align}

where $k$ is the density of the unnormalized exponential-logistic distribution:
\begin{equation}
    k = \int_{0}^{\infty} p(x) \, dx \quad k \in [0,\, 1] \label{eq-k}
\end{equation}

By definition, the density of the exponential-logistic distribution is less than 1, because the exponential PDF has density equal to 1 and it is multiplied by the logistic function, which lies between 0 and 1 (Eq.~\ref{eq-p-x}). Putting together Eqs.~\ref{eq-n_a}--\ref{eq-k}:

\begin{equation}
    \text{N}_\text{new} = \text{N}_\text{obs} \left(\frac{1-k}{k}\right)
\end{equation}

The combination of the new sample with the old observations forms a dataset that mimic the distribution of the true, unbiased population (Eq.~\ref{eq-p-x-unbias}) while maintaining the integrity of the original data.

\subsection{Results and discussion}

\begin{figure}[tb]
    \includegraphics[scale=0.9, keepaspectratio]{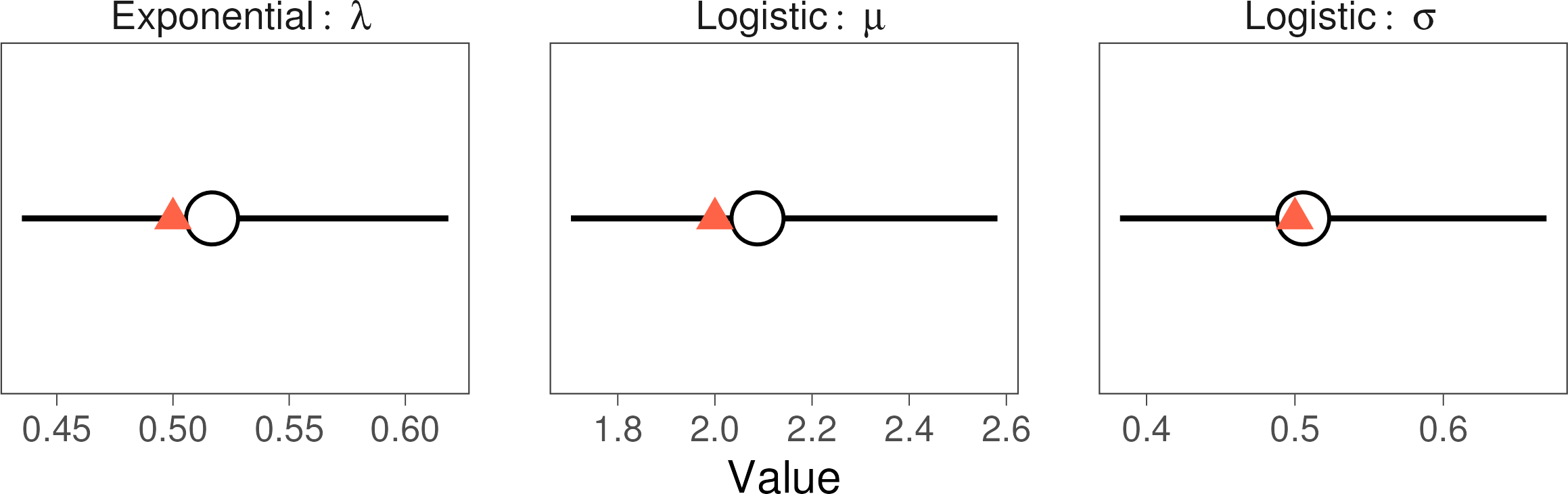}
        \centering
        \caption{Median (circle) and 95\% percentile interval (PI; horizontal bar) for the parameter estimate. The true value of each parameter (triangle) is contained in the interval, suggesting that the inference is accurate.}
        \label{fig-results}
\end{figure}

\begin{figure}[tb]
    \includegraphics[scale=0.9, keepaspectratio]{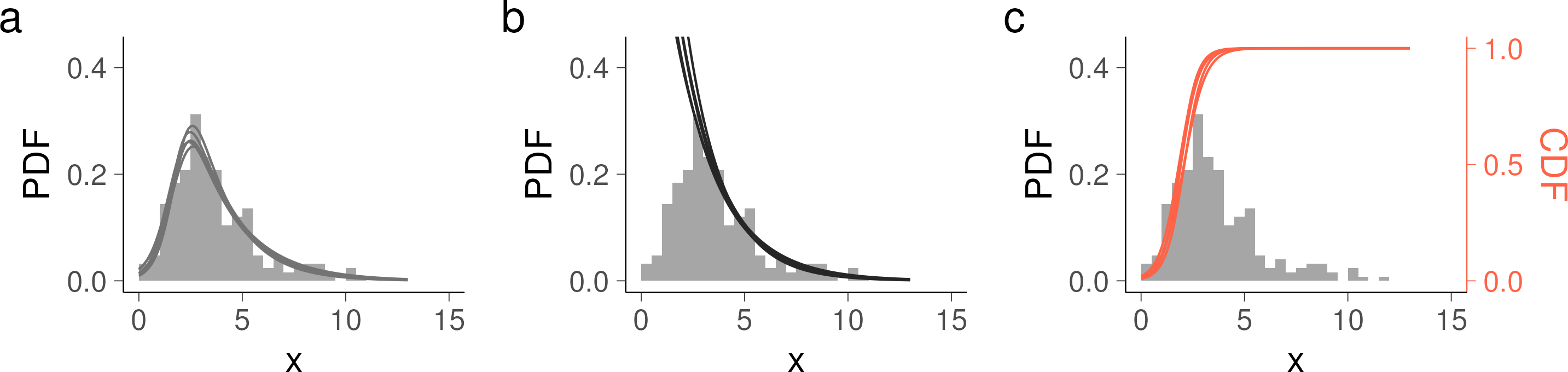}
        \centering
        \caption{Posterior predictive check. A set of plausible probability functions from the posterior distribution are overlaid on the histogram of the observed data. (a) Probability density function (PDF) for the exponential-logistic distribution (gray); (b) PDF for the exponential distribution (black); (c) Cumulative distribution function (CDF) for the logistic distribution (logistic function in red).}
        \label{fig-ppc}
\end{figure}

\begin{figure}[tb]
    \includegraphics[scale=0.9, keepaspectratio]{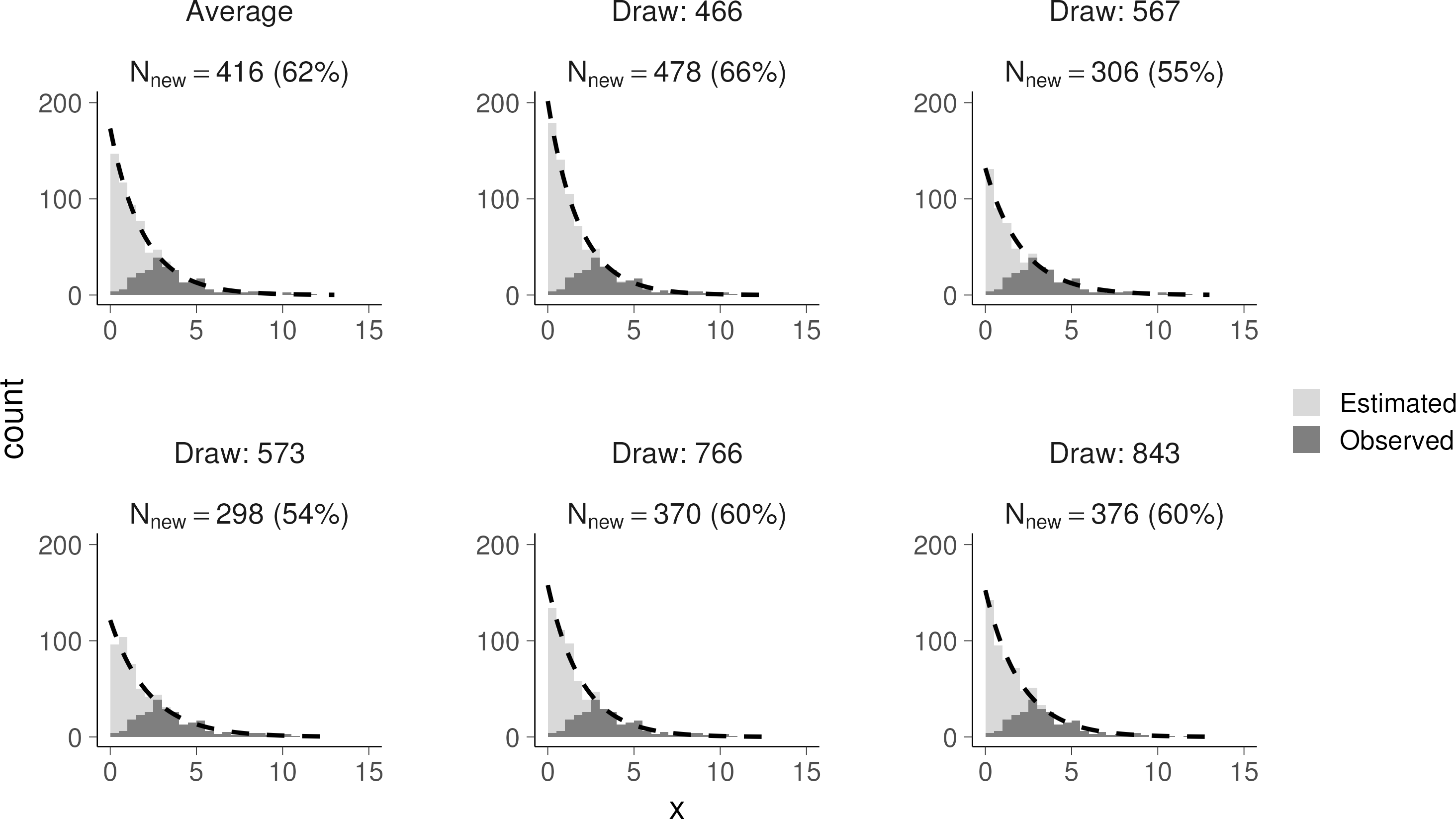}
        \centering
        \caption{Example of plausible complete samples. A new set of observations (light gray) was added to the original sample (dark gray) by drawing a combination of parameters from the posterior distribution. The black dashed line represents the estimated probability density function of the exponential distribution underlying the complete sample. The top left corner is the dataset composed using the mean of the parameter's value. }
        \label{fig-draws}
\end{figure}

The model recovered the true values of the parameters $\lambda$, $\mu$, and $\sigma$ (Eq.~\ref{eq-p-x}) that were used to generate the data. The true values are well within the 95\% percentile interval (PI) and close to the center of the marginal posterior distribution (Fig.~\ref{fig-results}). The combination of the new sample with the old observations forms a dataset that mimics the distribution of the true, unbiased population (Eq.~\ref{eq-p-x}) while maintaining the integrity of the original data. The model is validated by the posterior predictive check (Fig.~\ref{fig-ppc}). The output of the Bayesian regression is a distribution over the parameter's value (posterior distribution) rather than a single point estimate, which improves robustness to uncertainty; Fig.~\ref{fig-ppc} shows multiple curves that are all plausible, given the data and the modeling choices.

Once the values of the parameters of the exponential-logistic distribution (Eq.~\ref{eq-p-x}) have been established, the parameters of the exponential distribution (population distribution) and the logistic distribution (sampling mechanism) are also known. These parameters are used to recover the missing data. Fig.~\ref{fig-draws}, for example, shows $N_\text{draws} = 5$ plausible datasets created from a corresponding number of draws for $\lambda$, $\mu$, and $\sigma$ from the posterior distribution. All plausible datasets are identical for the observed data, but they differ in the imputed values. Given $N_\text{obs} = 250$, between 229--1111 new observations are needed to complete the dataset (Eq.~\ref{eq-n_b}), which corresponds to a percentage of missingness of 18\%–52\%.

Sampling variation is particularly beneficial when combined with multiple imputation techniques (\cite{lubbeInjuryRiskCurves2024}; \cite[ch. 2]{vanbuurenFlexibleImputationMissing2018}). Depending on the application, it may be more convenient to create a new single dataset based on the mean of the model parameters, instead of generating multiple replicas (see \textit{Average} dataset in Fig.~\ref{fig-draws}).

\section{Example with real-world data}
\subsection{Data}
The second example uses real-world data from CISS \autocite{nhtsa-ciss}, an open-access database that contains a representative sample of real-world crashes in the United States. This example is intended to be illustrative; there are nuances related to how crash databases are constructed that are ignored for simplicity. The CISS database includes detailed information on the vehicles and occupants involved. The data can be used to create injury risk curves (IRC), statistical models of occupants' injury outcome based on, for example, the severity of the crash. To estimate injury risk accurately, a complete baseline for cases with no injuries is needed (see conditional vs. unconditional IRC in \cite{lubbeInjuryRiskCurves2024}). A complete baseline, however, is missing in CISS because of the bias toward high-severity crashes.

CISS data from $2022$ were downloaded from the NTHSA repository \autocite{nhtsa-ciss-download}. The tables \verb|GV| (containing general information about the vehicles) and \verb|OCC| (containing information on the occupants) included in CISS were combined \autocite{Radja2022}. \textit{Reported} rear-end personal damage-only (PDO) crashes were selected. PDO crashes are those in which none of the occupants in any of the vehicles involved were injured. Within each crash record (\verb|CASEID|), which contains information on all vehicles involved, the following inclusion criteria were true:

\begin{itemize}
    \item \verb|MAIS == 0| to select uninjured cases;
    \item \verb|BODYCAT IN (1..6)| to select passenger vehicles; 
    \item \verb|CRASHCONF == "D"| to select rear-end crashes.
\end{itemize}

where \verb|MAIS| is the standard measure of injury outcome (\textit{Maximum Abbreviated Injury Scale} \cite{aaam-mais}). Then, we retained the elements with $\Delta v$ available (\verb|DVTOTAL != 999|; the absolute change in vehicle velocity at the time of impact is a common measure of crash severity).

The filtering identified 572 PDO vehicles, representing approximately $1.2$ million PDO vehicles after weighting.

\subsection{Method}
The analysis applies the procedure explained in Sec.~\ref{sec-example-simulated}. However, the \textit{brms} formula was slightly modified to accepts weighted data. All crashes in CISS are assigned a weight \verb|CASEWGT|, which is used to represent all police reported motor vehicle crashes occurring in the United States in that year \autocite[][Ch.~9]{zhangCrashInvestigationSamplinga}. The model formula was set to \verb"DVTOTAL | weights(CASEWGT) ~ 1". When resampling, the weight of each new data point was set to 1.

The decision to use an exponential distribution and a logistic function (Eq.~\ref{eq-p-x}) was inspired by \textcite{andricevicInjuryRiskFunctions2018} and \textcite{bargmanMethodologicalChallengesScenario2024}, who applied similar approaches in related contexts. In general, the exponential distribution is a good candidate because it concentrates its mass on low-speed impacts. While there could be other valid combinations of distributions, we have chosen this particular pairing for the sake of simplicity and consistency with previous literature, making it easier to illustrate the core concepts in this example.

Even when working with real data, it is good practice to do model checking with simulated data under controlled conditions before proceeding with the real data. Model-checking steps include constructing the model, simulating data from it, and checking that the model can infer the set parameters. This process ensures that the model and the inference engine are correctly specified \autocite[ch.~7]{gelmanRegressionOtherStories2020a}.

\subsection{Results and discussion}

\begin{figure}[tb]
    \includegraphics[scale=0.9, keepaspectratio]{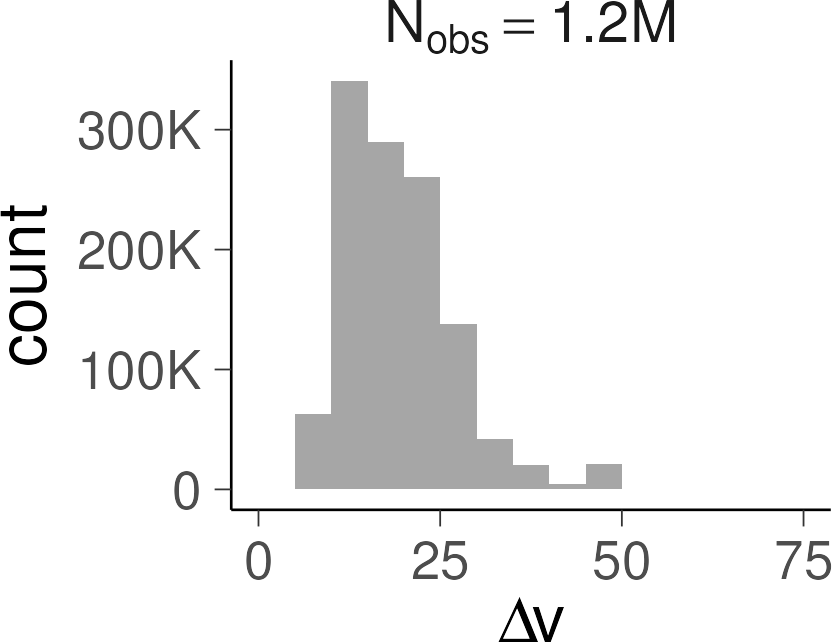}
        \centering
        \caption{The distribution of $\Delta v$ for the reported uninjured cases in the \textit{Crash Investigation Sampling System}. The histogram is weighted.}
        \label{fig-nhtsa-obs}
\end{figure}

\begin{figure}[tb]
    \includegraphics[scale=0.9, keepaspectratio]{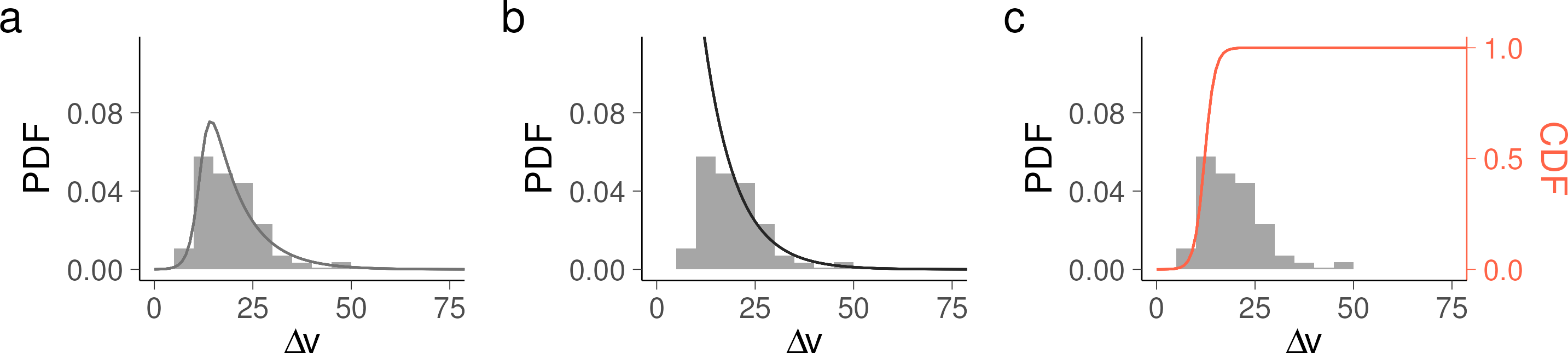}
        \centering
        \caption{Posterior predictive check. A set of plausible probability functions from the posterior distribution are overlaid on the weighted histogram of the observed data. Because the inference has low uncertainty, most lines overlap; (a) Probability density function (PDF) for the exponential-logistic distribution (gray); (b) PDF for the exponential distribution (black); (c) Cumulative distribution function (CDF) for the logistic distribution (logistic function in red).}
        \label{fig-nhtsa-ppc}
\end{figure}

\begin{table}[tb]
    \centering
    \caption{Summary of the posterior distribution for the parameters from the model, including median values and 95\% percentile intervals (PI).}
    \label{tab-nhtsa-params}
    \begin{tabular}{lcc}
    \toprule
    & Median & 95\% PI \\
    \midrule
    $\lambda$           & 0.12   & 0.12--0.12 \\
    $\mu_\text{bias}$    & 12.2  & 12.2--12.2 \\
    $\sigma_\text{bias}$ & 1.30   & 1.29--1.30 \\
    \bottomrule

    \end{tabular}
\end{table}

\begin{figure}[tb]
    \includegraphics[scale=0.9, keepaspectratio]{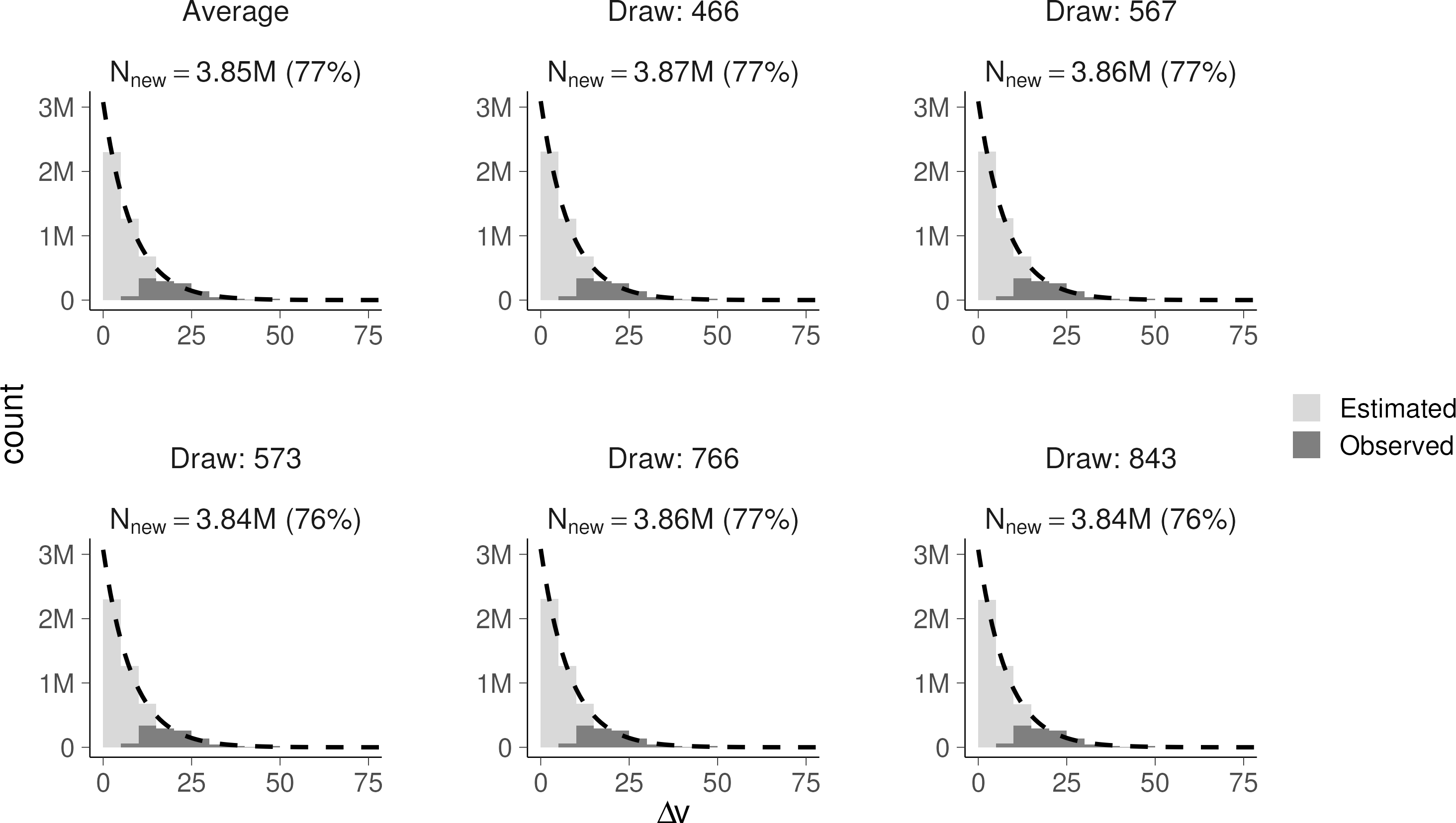}
        \centering
        \caption{Example of plausible complete datasets. A new set of observations (light gray) was added to the original weighted sample (dark gray) by drawing a combination of parameters from the posterior distribution. The black dashed line represents the estimated probability density function of the exponential distribution underlying the complete sample. The top left corner is the dataset composed using the mean of the parameters value.}
        \label{fig-nhtsa-draws}
\end{figure}

The distribution of $\Delta v$ for PDO crashes resembles the one shown in the simulated data example, where data appear to be missing in the left tail (Fig.~\ref{fig-nhtsa-obs}). Also in this case, the model can mimic the observations well (Fig.~\ref{fig-nhtsa-ppc}) and it supports the idea that data on PDO vehicles may be reported in a way that is compatible with an exponential-logistic distribution. A summary of the estimated parameters $\lambda$, $\mu$, and $\sigma$ from Eq.~(\ref{eq-p-x}) is given in Tab.~\ref{tab-nhtsa-params} (the uncertainty intervals are narrow because of the large quantity of data). Examples of plausible complete datasets recovered from posterior distribution draws are shown in \ref{fig-nhtsa-draws}. As with the previous example, all plausible datasets are identical for the observed data, but the imputed values differ. About $3.8$ million new observations were imputed, which corresponds to about 77\% missing data.

Rear-end collisions are frequent, accounting for about 29\% of all reported crashes involving PDO vehicles \autocite[][Tab.~29]{nhtsa2024traffic}. Additionally, passenger vehicles---cars, vans, and light trucks---constitute approximately 81\% of all reported PDO vehicles \autocite[][Tab.~36]{nhtsa2024traffic}. Given that many rear-end collisions result in minor bumper-to-bumper impacts, vehicle owners often decide for private repairs or choose to ignore the damage, leading to higher underreporting rates. Therefore, it is plausible that the underreporting percentage of PDO passenger vehicles involved in rear-end collisions could exceed the general 60\% estimate by \textcite{blincoe2023economic}.

Our estimate of a 77\% underreporting rate appears reasonable and it aligns with broader underreporting trends. The corresponding $3.8$ million imputed cases is reasonably higher than the $2.2$ million \textit{police-reported} PDO passenger vehicles involved in rear-ends in 2022, which are also prone to undercounting \autocite{nhtsa-first-query}. Unfortunately, cross-study comparisons and validation are challenging due to the lack of public data. If the distribution of missing $\Delta v$ in PDO rear-end crashes mirrors that of observed data, the method proposed here should be valid. However, this figure should be used cautiously until it is systematically validated.

Compared to previous studies, our approach has some advantages:

\begin{enumerate}
    \item It can be applied to datasets of any size, from small to large, although more informative priors would be required when the number of observations is limited;
    \item It does not depend on external data, which are often not publicly available (e.g., insurance data), but external data is necessary for validation;
    \item It is applied to the same data that will be used for further analyses, which ensures consistency between observations and imputed values;
    \item It retrieves the proportion of missing data but also the distribution of crash measures (e.g., $\Delta v$), which is key for injury risk and safety benefit estimation;
    \item It is based on probability principles without relying on arbitrary decisions for model fitting, which enhances adaptability, clarity, and reliability;
    \item It yields a comprehensive account of inference uncertainty, which improves robustness when coupled with multiple-data imputation techniques;
    \item It can be extended to accommodate data with hierarchical structures (multilevel models for vehicle, enviroment, and occupant features).
\end{enumerate}

\section{Conclusions}
Sampling bias that leads to MNAR data can be challenging to identify and address. Classical approaches to imputing missing data based on the qualities of the available data are ineffective, as samples affected by MNAR do not contain enough information to recover the missing portion. Effectively addressing this bias requires domain expertise and a thorough understanding of the data collection process. Our approach leverages the Bayesian framework to encode domain expertise and prior knowledge into a statistical model that is transparent and handles uncertainties intuitively.

The method was applied to a crash database, which tends to overrepresent high-severity crashes due to specific reporting criteria. We developed a model that combines an exponential distribution with a logistic function to explain the crash reporting mechanism. The results are compatible with previous, albeit sparse, findings demonstrating the potential of our approach. This method can be adapted for other use cases in different domains, to construct models for sampling processes governed by different probability functions.

\section*{Acknowledgements}
I thank my colleagues at Autoliv Research for their comments and suggestions on this paper; the members of the Stan forum (\url{https://discourse.mc-stan.org}) for their technical support; and Tina Mayberry for the language review.

\section*{Disclosure of interest}
A. Morando is employed at Autoliv (\url{www.autoliv.com}), which is a company that develops, manufactures, and sells (among others) vehicle safety products and personal protective equipment. Any opinions and recommendations expressed in this paper are those of the author and do not necessarily reflect the views of Autoliv.

\printbibliography

@online{CustomFamiliesBrms,
  title = {Custom {{Families}} in Brms {{Models}} — Custom\_family},
  author = {Bürkner, Paul-Christian},
  url = {https://paulbuerkner.com/brms/reference/custom_family.html},
  date = {2024},
  urldate = {2024-11-08},
  langid = {english}
}

@online{brmBrms,
  title = {Fit {Bayesian Generalized} (Non-)Linear Multivariate Multilevel Models},
  author = {Bürkner, Paul-Christian},
  url = {https://paulbuerkner.com/brms/reference/brm.html},
  date = {2024},
  urldate = {2024-11-08},
  langid = {english}
  }

@online{tidyverseTibble,
  title   = {Build a data frame},
  author  = {Kirill M\"uller and Hadley Wickham},
  url     = {https://tibble.tidyverse.org/reference/tibble.html},
  date    = {2025},
  urldate = {2025-02-04}
}

@incollection{allisonMissingData2009,
  title = {Missing Data},
  booktitle = {The {{Sage}} Handbook of Quantitative Methods in Psychology.},
  author = {Allison, Paul D.},
  date = {2009},
  pages = {72--89},
  publisher = {Sage Publications Ltd},
  location = {Thousand Oaks,  CA},
  doi = {10.4135/9780857020994.n4},
  isbn = {978-1-4129-3091-8 (Hardcover)}
}

@article{kruschkeBayesianDataAnalysis2017,
  title = {Bayesian Data Analysis for Newcomers},
  author = {Kruschke, John K. and Liddell, Torrin M.},
  date = {2017},
  journaltitle = {Psychonomic Bulletin \& Review},
  pages = {1--23},
  issn = {1531-5320},
  doi = {10.3758/s13423-017-1272-1}
}

@article{carpenterStanProbabilisticProgramming2017,
  title = {\textit{Stan} : A Probabilistic Programming Language},
  volume = {76},
  issn = {1548-7660},
  url = {http://www.jstatsoft.org/v76/i01/},
  doi = {10.18637/jss.v076.i01},
  shorttitle = {\textit{Stan}},
  number = {1},
  journaltitle = {Journal of Statistical Software},
  shortjournal = {J. Stat. Soft.},
  author = {Carpenter, Bob and Gelman, Andrew and Hoffman, Matthew D. and Lee, Daniel and Goodrich, Ben and Betancourt, Michael and Brubaker, Marcus and Guo, Jiqiang and Li, Peter and Riddell, Allen},
  urldate = {2024-11-11},
  date = {2017},
  langid = {english},
}

@book{vanbuurenFlexibleImputationMissing2018,
  location = {United Kingdom},
  edition = {2nd},
  title = {Flexible Imputation of Missing Data},
  isbn = {978-1-138-58831-8},
  url = {https://stefvanbuuren.name/fimd/},
  publisher = {Chapman and Hall/{CRC}},
  author = {van Buuren, Stef},
  urldate = {12-11-2024},
  date = {2018},
}

@online{mcstanUnboundedContinuousLogistic,
	author = {{Stan Development Team}},
	title = {Stan Functions Reference - {L}ogistic distribution},
	url = {https://mc-stan.org/docs/functions-reference/unbounded_continuous_distributions.html#logistic-distribution},
	year = {2024},
  urldate = {12-11-2024}
}

@online{mcstanUnboundedContinuousExponential,
	author = {{Stan Development Team}},
	title = {Stan Functions Reference - {E}xponential distribution},
	url = {https://mc-stan.org/docs/functions-reference/positive_continuous_distributions.html#exponential-distribution},
	year = {2024},
  urldate = {12-11-2024}
}

@book{gelmanRegressionOtherStories2020a,
  edition = {1},
  title = {Regression and Other Stories},
  isbn = {978-1-139-16187-9 978-1-107-02398-7 978-1-107-67651-0},
  url = {https://www.cambridge.org/highereducation/product/9781139161879/book},
  publisher = {Cambridge University Press},
  author = {Gelman, Andrew and Hill, Jennifer and Vehtari, Aki},
  urldate = {2022-02-01},
  date = {2020-07-23},
  doi = {10.1017/9781139161879},
}

@Manual{R,
    title = {R: A Language and Environment for Statistical Computing},
    author = {{R Core Team}},
    organization = {R Foundation for Statistical Computing},
    address = {Vienna, Austria},
    year = {2023},
    url = {https://www.R-project.org/},
  }

@Article{burknerBrmsPackageBayesian2017,
    title = {{brms}: An {R} Package for {Bayesian} Multilevel Models Using {Stan}},
    author = {Paul-Christian Bürkner},
    journal = {Journal of Statistical Software},
    year = {2017},
    volume = {80},
    number = {1},
    pages = {1--28},
    doi = {10.18637/jss.v080.i01},
    encoding = {UTF-8},
  }

@Misc{Stan,
    title = {{RStan}: the {R} interface to {Stan}},
	  author = {{Stan Development Team}},
    note = {R package version 2.32.6},
    year = {2024},
    url = {https://mc-stan.org/},
  }

@Manual{cmdstanr,
    title = {cmdstanr: R Interface to 'CmdStan'},
    author = {Jonah Gabry and Rok Češnovar and Andrew Johnson and Steve Bronder},
    year = {2024},
    note = {https://mc-stan.org/cmdstanr/, https://discourse.mc-stan.org},
  }

@online{integrand1d,
	author = {{Stan Development Team}},
	title = {Stan Functions Reference - {H}igher order functions},
	url = {https://mc-stan.org/docs/functions-reference/higher-order_functions.html#specifying-an-integrand-as-a-function},
	year = {2024},
  urldate = {12-11-2024}
}

@article{pathfinder2022,
  author  = {Lu Zhang and Bob Carpenter and Andrew Gelman and Aki Vehtari},
  title   = {Pathfinder:  Parallel quasi-Newton variational inference},
  journal = {Journal of Machine Learning Research},
  year    = {2022},
  volume  = {23},
  number  = {306},
  pages   = {1--49},
  url     = {http://jmlr.org/papers/v23/21-0889.html}
}

@Manual{rformula,
    title = {{Stats}: Model Formulae},
    author = {{R Core Team}},
    organization = {R Foundation for Statistical Computing},
    address = {Vienna, Austria},
    year = {2023},
    url = {https://stat.ethz.ch/R-manual/R-devel/library/stats/html/formula.html},
  }

@article{bargmanMethodologicalChallengesScenario2024,
  title = {Methodological challenges of scenario generation validation: A rear-end crash-causation model for virtual safety assessment},
  volume = {104},
  issn = {13698478},
  url = {https://linkinghub.elsevier.com/retrieve/pii/S1369847824000810},
  doi = {10.1016/j.trf.2024.04.007},
  shorttitle = {Methodological challenges of scenario generation validation},
  pages = {374--410},
  journaltitle = {Transportation Research Part F: Traffic Psychology and Behaviour},
  shortjournal = {Transportation Research Part F: Traffic Psychology and Behaviour},
  author = {B\"argman, Jonas and Svärd, Malin and Lundell, Simon and Hartelius, Erik},
  urldate = {2024-08-26},
  date = {2024-07},
  langid = {english},
}

@article{lubbeInjuryRiskCurves2024,
  title = {Injury risk curves to guide safe speed limits on Swedish roads using German crash data supplemented with estimated non-injury crashes},
  volume = {202},
  issn = {0001-4575},
  url = {https://www.sciencedirect.com/science/article/pii/S0001457524001313},
  doi = {10.1016/j.aap.2024.107586},
  pages = {107586},
  journaltitle = {Accident Analysis \& Prevention},
  shortjournal = {Accident Analysis \& Prevention},
  author = {Lubbe, Nils and Jeppsson, Hanna and Sternlund, Simon and Morando, Alberto},
  urldate = {2024-11-26},
  date = {2024-07-01},
}

@article{andricevicInjuryRiskFunctions2018,
  title = {Injury risk functions for frontal oblique collisions},
  volume = {19},
  issn = {1538-9588, 1538-957X},
  url = {https://www.tandfonline.com/doi/full/10.1080/15389588.2018.1442926},
  doi = {10.1080/15389588.2018.1442926},
  pages = {518--522},
  number = {5},
  journaltitle = {Traffic Injury Prevention},
  shortjournal = {Traffic Injury Prevention},
  author = {Andricevic, Nino and Junge, Mirko and Krampe, Jonas},
  urldate = {2022-02-10},
  date = {2018-07-04},
  langid = {english},
}

@techreport{blincoe2023economic,
  author       = {Blincoe, L. and Miller, T. and Wang, J.-S. and Swedler, D. and Coughlin, T. and Lawrence, B. and Guo, F. and Klauer, S. and Dingus, T.},
  title        = {The Economic and Societal Impact of Motor Vehicle Crashes, 2019 (Revised)},
  year         = {2023},
  month        = {2},
  number       = {DOT HS 813 403},
  institution  = {{National Highway Traffic Safety Administration}},
  type         = {Report}
}

@online{nhtsa-ciss,
  author  = {{National Highway Traffic Safety Administration},},
  shortauthor = {{NHTSA}},
  title   = {{Crash Investigation Sampling System}},
  url     = {https://www.nhtsa.gov/crash-data-systems/crash-investigation-sampling-system},
  year    = {2025},
  urldate = {2025-02-04}
}

@online{aaam-mais,
  author  = {{Association for the Advancement of Automotive Medicine},},
  shortauthor = {{AAAM}},
  title   = {{Abbreviated Injury Scale (AIS)}},
  url     = {https://www.aaam.org/abbreviated-injury-scale-ais-position-statement/},
  year    = {2025},
  urldate = {2025-02-04}
}

@online{nhtsa-ciss-download,
  author  = {{National Highway Traffic Safety Administration},},
  shortauthor = {NHTSA},
  title   = {{NHTSA File Downloads}},
  url     = {https://www.nhtsa.gov/file-downloads?p=nhtsa/downloads/CISS/},
  year    = {2025},
  urldate = {2025-02-04}
}

@techreport{Radja2022,
  author      = {Radja, G. A. and Noh, E.-Y. and Zhang, F.},
  title       = {{Crash Investigation Sampling System -- 2021 analytical user's manual}},
  year        = {2022},
  number      = {DOT HS 813 398},
  institution = {{National Highway Traffic Safety Administration}},
  type        = {Report},
  url         = {https://crashstats.nhtsa.dot.gov/Api/Public/Publication/813398},
  urldate     = {2025-02-04}
}

@techreport{zhangCrashInvestigationSamplinga,
  title       = {Crash {{Investigation Sampling System}}: {{Sample Design}} and {{Weighting}}},
  author      = {Zhang, Fan and Noh, Eun Young and Subramanian, Rajesh and Chen, Chou-Lin},
  langid      = {english},
  date        = {2019},
  number      = {DOT HS 812 804},
  institution = {National Highway Traffic Safety Administration}
}

@techreport{nhtsa2024traffic,
  author      = {{National Center for Statistics and Analysis}},
  shortauthor = {{NCSA}},
  title       = {Traffic Safety Facts 2022: A Compilation of Motor Vehicle Traffic Crash Data},
  institution = {{National Highway Traffic Safety Administration}},
  year        = {2024},
  month       = {12},
  number      = {DOT HS 813 656},
  url         = {https://crashstats.nhtsa.dot.gov/Api/Public/ViewPublication/813656.pdf}
}

@online{nhtsa-first-query,
  author      = {{National Highway Traffic Safety Administration},},
  shortauthor = {NHTSA},
  title       = {{Fatality and Injury Reporting System Tool (FIRST)}},
  url         = {https://cdan.dot.gov/query},
  year        = {2025},
  urldate     = {2025-03-04}
}

\appendix

\section{Some code implementation details}
The simulated dataset was created as a \verb|tidyverse::tibble| \autocite{tidyverseTibble}:

\begin{lstlisting}[caption={Creation of the dataframe containing the simulated data}, label={code-data-tibble}]
df <- tibble(x = rexp(2500, rate = 0.5)) %>%
    mutate(w = plogis(x, location = 2.0, scale = 0.5)) %>%
    slice_sample(n = 250, weight_by=w)

% > df
% # A tibble: 250 x 2
%       x      w
%   <dbl>  <dbl>
%  1 1.83  0.413
%  2 3.18  0.914
%  3 2.92  0.863
%  4 2.20  0.600
%  5 3.23  0.922
%  6 4.45  0.993
%  7 1.44  0.245
%  8 3.32  0.934
%  9 6.13  1.000
% 10 0.95  0.110
% # i 240 more rows
\end{lstlisting}

The custom exponential-logistic distribution can be created with \verb|brms::custom_family| \autocite{CustomFamiliesBrms}. The function expects one of the parameters in \verb|dpars| to be named \verb|mu|, the parameter that the regression formula of the model will refer to \autocite{CustomFamiliesBrms}. We selected the exponential component of Eq.~(\ref{eq-p-x}) as the principal one, because it yields the population parameter from which we can eventually retrieve an unbiased sample. This choice is convenient because the mean of the exponential distribution is $\mu = \frac{1}{\lambda}$. The other arguments are related to the type of distribution, the link functions, and the domain bounds for all parameters in the custom family (the names for the other parameters are arbitrary):

\begin{lstlisting}[caption={Definition of the custom distribution family in brms.}, label={code-brms-custom-family}]
explogi <- brms::custom_family(
    name = "explogi",
    type = "real",
    dpars = c("mu", "muBias", "sigmaBias"),
    lb = c(0, NA, 0),
    links = c("log", "identity", "log"),
)
\end{lstlisting}

Once the properties of the custom distribution are defined in \textit{brms}, we need to write the \textit{Stan} code that computes the custom probability function. The function that defines the PDF in \textit{Stan} needs to be called \verb|<name>_lpdf|, where \verb|<name>| is the name given to \verb|brms::custom_family| (in this case \verb|explogi|). The suffix \verb|_lpdf| indicates that the computation is done in the log domain (i.e., the multiplication of probabilities becomes addition) for precision and performance reasons. Conveniently, we can compose the exponential-logistic function from the built-in \verb|exponential_lpdf| \autocite{mcstanUnboundedContinuousExponential} and \verb|logistic_lcdf| \autocite{mcstanUnboundedContinuousLogistic}. Then, the resulting mix of distributions need to be normalized to a valid PDF (i.e., integrated to 1):

\begin{lstlisting}[caption={Definition of the custom log probability density function in Stan.}, label={code-stan-lpdf}]
stan_lpdf <- "
    real explogi_lpdf(
        real x,
        real mu,
        real muBias,
        real sigmaBias) {

        real log_exp_pdf = exponential_lpdf(x | 1/mu);
        real log_logistic_cdf = logistic_lcdf(x | muBias, sigmaBias);

        real log_normalizer = log(integrate_1d(
                                    integrand,
                                    0.0, positive_infinity(),
                                    {mu, muBias, sigmaBias},
                                    {0}, {0}, 1e-06));

        return log_exp_pdf + log_logistic_cdf - log_normalizer;
  }
"
\end{lstlisting}

The normalization factor \verb|log_normalizer| is computed by integration. This approach is the most general and flexible, although it makes sampling slower. The alternative is to find a close-form solution, but this possibility exists only for some specific mix of distributions. The integration requires the integrand, the limit of integration, and the parameters to all be passed to the integrand. The integrand follows a special signature \autocite{integrand1d} and contains the code to compute the exponential-logistic distribution to be integrated (in this case, in the linear domain):

\begin{lstlisting}[caption={Definition of the integrand function in Stan.}, label={code-stan-integrand}]
stan_integrand <- "
    real integrand(
        real x,
        real xc,
        array[] real theta,
        data array[] real x_r,
        data array[] int x_i) {

        real mu = theta[1];
        real muBias = theta[2];
        real sigmaBias = theta[3];

        real exp_pdf = exp(exponential_lpdf(x | 1/mu));
        real logistic_cdf = exp(logistic_lcdf(x | muBias, sigmaBias));

        return exp_pdf * logistic_cdf;
}
"
\end{lstlisting}

Finally, the \textit{Stan} functions, stored in the \verb|stanvars| variable, can be supplied to \textit{brms} \autocite{CustomFamiliesBrms}:

\begin{lstlisting}[caption={Definition of the variable containing the Stan code.}, label={code-stan-fun}]
stan_funs <- paste(stan_lpdf, stan_integrand, sep = "\n")
stanvars <- stanvar(scode = stan_funs, block = "functions")
\end{lstlisting}

The functions to compute the exponential-logistic distribution (Eq.~\ref{eq-p-x}) are now defined and can be used for modeling with the command \verb|brms::brm| \autocite{brmBrms}:

\begin{lstlisting}[caption={Definition of brms model and sampling settings.}, label={code-brm}]
m <- brm(
    bf(x ~ 1),
    data = df,
    family = explogi,
    stanvars = stanvars,
    prior = c(
        prior(normal(0, 5), class = "Intercept"),
        prior(normal(0, 5), class = "muBias"),
        prior(normal(0, 1), class = "sigmaBias")
    ),
    algorithm = "pathfinder",
    seed = 23,
    file = "m.rds",
    file_refit = "on_change"
)
\end{lstlisting}

The formula \verb|x ~ 1| is the standard syntax to define a regression model in \textit{R} \autocite{rformula}; it indicates that the model predicts the distribution of the observed variable \verb|x| in the dataset \verb|df| (Lst.~\ref{code-data-tibble}) by modelling the \verb|mu| of the exponential-logistic distribution ($\mu = \frac{1}{\lambda}$). In this case, we do not have predictors, so the model has only the coefficient $\alpha$ for the intercept, which is estimated in the log space according to the specified link function:

\begin{equation}
    \begin{aligned}
        x & \sim \text{Exponential-Logistic}(\mu, \mu_{\text{bias}}, \sigma_{\text{bias}}) \\
        log(\mu) &= \alpha
    \end{aligned}
    \label{eq-link-fun}
\end{equation}

The remaining family parameters (\verb|muBias| and \verb|sigmaBias|) are estimated automatically in a similar manner (intercept only); they do not need to be explicitly declared in the formula.

\end{document}